\documentclass[a4paper,11pt]{article}
\pdfoutput=1 

\usepackage{jcappub} 

\usepackage[T1]{fontenc} 

\usepackage{amssymb}
\usepackage{amsmath, graphicx}
\usepackage{slashed}
\usepackage{csquotes}
\usepackage{braket}
\usepackage{natbib}
\usepackage{array}
\usepackage{xcolor}

\newcommand\bea{\begin{eqnarray}}
\newcommand\eea{\end{eqnarray}}
\usepackage{amsmath}
\usepackage{url}
\usepackage{natbib}
\usepackage{aas_macros}

\usepackage{cancel}

\newcommand{\be}{\begin{equation}}
\newcommand{\ee}{\end{equation}}

\newcommand{\beq}{\begin{eqnarray}}
\newcommand{\eeq}{\end{eqnarray}}

\title{\boldmath Gravitational wave signatures of primordial black hole accretion during early matter domination}

\author[a]{Rouzbeh Allahverdi, }
\emailAdd{rouzbeh@unm.edu}

\author[b]{James B. Dent, }
\emailAdd{jbdent@shsu.edu}

\author[a]{Ngo Phuc Duc Loc, }
\emailAdd{locngo148@gmail.com}

\author[c]{and Tao Xu}
\emailAdd{tao.xu@ou.edu}

\affiliation[a]{Department of Physics and Astronomy, University of New Mexico, Albuquerque, NM 87131, USA}
\affiliation[b]{Department of Physics, Sam Houston State University, Huntsville, TX 77341, USA}
\affiliation[c]{Homer L. Dodge Department of Physics and Astronomy, University of Oklahoma, Norman, OK 73019, USA}

\abstract{
We present a scenario in which primordial black holes (PBHs) form in a post-inflationary radiation-dominated (RD) phase and then experience significant accretion during a phase of early matter dominated (EMD). We show that PBH masses could grow by up to two orders of magnitude. Restricting to the linear perturbation regime, we compute the gravitational wave (GW) spectrum that features two peaks. The high-frequency peak is associated with the PBH formation in the RD phase, while the low-frequency peak is due to the sudden transition from EMD to the later, standard RD phase. We identify a PBH mass range where one or both peaks can be observed by a combination of different GW detectors. Finally, we show the signal-to-noise ratio of the total GW spectrum for PBHs in the asteroid mass window, where they could comprise the totality of dark matter.
}

\begin{document}

\hfill{\small CETUP-2024-011}

\maketitle
\flushbottom

\section{Introduction}
\label{sec:introduction}

The study of primordial black holes (PBHs), black holes formed not through standard astrophysical processes but rather through mechanisms in the early universe, goes back several decades, dating to the pioneering works found in~\cite{Zeldovich:1967lct,Hawking:1971ei,Carr:1974nx,Carr:1975qj}. PBH formation mechanisms include the collapse of a variety of possible entities including: large curvature perturbations produced during inflation~\cite{Ivanov:1997ia,Garcia-Bellido:1996mdl,Yokoyama:1995ex,Kawasaki:1997ju,Garcia-Bellido:2017mdw,Ballesteros:2017fsr}, cosmic domain walls~\cite{Liu:2019lul}, condensates~\cite{Kawana:2021tde} or bubble walls formed during a first-order phase transition (FOPT) in the early universe~\cite{Deng:2017uwc,Flores:2024lng}, and solitons~\cite{Cotner:2019ykd}. For reviews, including an overview of some formation mechanisms and constraints on the allowed abundance of PBHs, see for example~\cite{Carr:2020gox,Carr:2020xqk,Green:2020jor,Escriva:2022duf} (for an observationally motivated case for PBHs, see~\cite{Carr:2023tpt}). 

Since the formation of PBHs could be connected to early universe cosmology scenarios with additional ramifications for beyond the standard model (BSM) particle physics (in the cases of PBH formation via domain walls, in conjunction with a FOPT, or from soliton collapse, for example), the observation of a PBH would be a momentous achievement. Along with these exciting possibilities, which could provide a window into the early universe and BSM sectors, yet another intriguing aspect of PBHs is that they could provide some or all of the cold dark matter (DM) content in the universe. The possibility of PBH being DM has long been considered, with a variety of constraints regarding the fraction of DM that can currently exist (or could have existed at the time of PBH formation) as DM. An obvious and necessary property is that they need to have not evaporated by the present day, which sets a lower bound on the mass of $~5\times10^{14}$~g\footnote{Lighter PBHs could have evaporated in the early universe and non-thermally produced the DM population~\cite{Allahverdi:2017sks,Morrison:2018xla, Gehrman:2023qjn}.}. There is currently an open region of parameter space dubbed `the asteroid-mass window' in which PBHs could form the totality of DM\footnote{This mass range was opened after several constraints in this region of parameter space from lensing, dynamical capture in stars, and destruction of white dwarfs were revisited in~\cite{Katz:2018zrn,Montero-Camacho:2019jte}. However, more recently a study appeared that again places constraints in the asteroid-mass window from stellar capture in ultra-faint dwarf galaxies~\cite{Esser:2025pnt}.}. This window lies roughly in the mass range of $\sim 10^{17} - 10^{22}$~g, with bounds on the low-mass side from Hawking evaporation constraints~\cite{Coogan:2020tuf,Ray:2021mxu,DelaTorreLuque:2024qms,Tan:2024nbx} (see also~\cite{Auffinger:2022khh} for an overview and compendium of Hawking evaporation constraints) and on the high-mass side from gravitational lensing observations~\cite{Niikura:2017zjd,Montero-Camacho:2019jte,Smyth:2019whb,Croon:2020ouk,Mroz:2024wia}. Other possibilities for accessing a PBH population in this mass range include solar system dynamics~\cite{Tran:2023jci,Bhalla:2024jbu,Thoss:2024dkg}, future lensing programs~\cite{Bird:2022wvk,Tamta:2024pow,Fedderke:2024wpy}, and superradiant instabilities~\cite{Branco:2023frw,Dent:2024yje}.

An additional avenue for PBH exploration opened with the observation of gravitational waves (GWs) by the LIGO and Virgo collaborations~\cite{LIGOScientific:2016aoc}, which has continued apace with subsequent GW measurements from the LIGO and Virgo~\cite{LIGOScientific:2018mvr,LIGOScientific:2020ibl}, and the LIGO, Virgo, and KAGRA collaborations (LVK)~\cite{KAGRA:2021vkt}. The mass range of inspiraling binary black holes accessible to the LVK collaboration is roughly in the solar mass and above range (which corresponds to frequencies observable by LVK in the $\sim10~{\rm Hz}-{\rm few}~{\rm kHz}$ window), well outside the asteroid-mass space, though the possibility still exists that populations of PBHs could be the progenitors of some fraction of the observed GW signals (see~\cite{Carr:2019kxo}, for an example of a scenario predicting appreciable PBH populations within reach of the LVK observatories). Future planned ground-based observatories such as Cosmic Explorer~\cite{Reitze:2019iox} and Einstein telescope~\cite{Punturo:2010zz} will have more sensitivity than the LVK collaborations, but in the same relative frequency band (with NEMO~\cite{Ackley:2020atn} the possibility exists for a few kHz band ground-based observatory, which is the current upper edge of the frequency range for proposed observed observatories). Other current and future GW observatories are either in the same range as the LVK collaboration or well below (down to the nHZ range currently being probed by pulsar-timing array GW detectors such as NANOGrav~\cite{NANOGrav:2023gor}, EPTA~\cite{EPTA:2023fyk}, IPTA~\cite{Antoniadis:2022pcn}, 
 and CPTA~\cite{Xu:2023wog}, along with future astrometry based possibilities such as SKA~\cite{Janssen:2014dka} and THEIA~\cite{Garcia-Bellido:2021zgu}, with upcoming space-based GW observatories such as LISA~\cite{LISA:2017pwj,LISACosmologyWorkingGroup:2023njw}, TianQin~\cite{TianQin:2020hid}, and Taiji~\cite{Hu:2017mde} probing the mHz band, as well as BBO~\cite{Corbin:2005ny} and DECIGO~\cite{Kawamura:2020pcg}
 operating around dHz. The idea of a $\mu$Hz range observatory has also been floated in $\mu$-ARES~\cite{Sesana:2019vho}).

Although binary inspirals of PBHs in the asteroid-mass window are currently well beyond existing technological capabilities (they are in the $\gtrsim 10^{14}$~Hz range; see~\cite{Aggarwal:2020olq,Aggarwal:2025noe} for examples of ideas for probing high-frequency GWs), other GW signatures associated with PBHs might be within reach of some of the current and planned observatories discussed above. For example, scalar perturbations will source tensor modes at second order, which subsequently form a stochastic gravitational wave background (the study of such a scalar-induced gravitational wave (SIGW) signature goes back decades~\cite{Tomita:1967wkp,Ananda:2006af,Baumann:2007zm,Assadullahi:2009jc,Kohri:2018awv}. For a review see~\cite{Domenech:2021ztg}.) Some recent work has focused on the effect on a SIGW signal from an epoch of non-standard cosmology such as a period of early matter domination (EMD)~\cite{Fernandez:2023ddy,Kumar:2024hsi}. An EMD era on its own presents a set of rich possibilities for investigation including the alteration of the standard DM freeze-out story~\cite{Drees:2018dsj,Allahverdi:2018aux,Allahverdi:2019jsc}, or new features of structure formation including mini-clusters and mini-halos~\cite{Erickcek:2011us,Barenboim:2013gya,Fan:2014zua,Erickcek:2015jza,Delos:2018ueo,Delos:2021rqs,Shen:2022ltx}\footnote{Another scenario where PBHs produce GWs at second order is that of an early PBH dominated universe where isocurvature fluctuations from the spatially inhomogeneous distribution of PBHs source GWs. This occurs for PBH populations that are light enough to have decayed, reheating the universe, prior to the big bang nucleosynthesis (BBN) era~\cite{Papanikolaou:2020qtd,Domenech:2020ssp,Paul:2025kdd}.} (for reviews and discussions of EMD cosmology, see~\cite{Allahverdi:2020bys,Batell:2024dsi}).

It has been demonstrated that a determining factor in the amplitude of the SIGW signal associated with the EMD epoch is the rate of the transition from EMD to the standard radiation dominated (RD). Intriguingly, a sudden transition (where the transition occurs much faster than the Hubble time at that era) can create a dramatic amplification of the GW signal~\cite{Inomata:2019ivs,Domenech:2024wao}, rendering it possibly observable in near-term GW observatories (for SIGW in the context of a gradual transition, see~\cite{Inomata:2019zqy}, while works examining a range of transition timescales include~\cite{Pearce:2023kxp,Pearce:2025ywc}).

In the present work we examine the possibility of a novel GW profile produced by two separate factors: i) SIGW from scalar perturbations linked with the overdensities that create a PBH population prior to a period of EMD and ii) the effect on GWs due to the transition from EMD to RD. The combination of these two factors can create a double-peaked signature observable by either a single GW observatory or a combination of observatories.\footnote{For other processes that can produce a double-peaked feature in the GW spectrum see, for example, the works~\cite{Buen-Abad:2023hex,Dent:2024bhi} which examine GW production in phase transitions.} The PBH mass range under consideration falls within the asteroid-mass window, which provides the further opportunity for correlated signatures from future Hawking evaporation or gravitational lensing observations. Previous work~\cite{Inomata:2019ivs,Pearce:2023kxp,Pearce:2025ywc} has focused on SIGW from scalar perturbations of the nearly scale-invariant type consistent with the amplitude and spectral index determined through cosmic microwave background (CMB) measurements. As we are interested in perturbations large enough to collapse to a PBH population, we will investigate SIGW from a scalar perturbation spectrum that is substantially larger than those of the CMB-type. This is observationally allowed since the scales we are concerned with are much smaller than those measured via the CMB, where only $\Delta N_{\rm eff}$ constraints apply to the total energy density of GWs \cite{Smith:2006nka, Kohri:2018awv}.

Although PBH formation during an era of EMD~\cite{Khlopov:1980mg,Polnarev:1981,Green:1997pr,Harada:2016mhb} can differ significantly from formation during RD\footnote{For example, due to the reduction in pressure from the cosmological background, enhanced formation rates are expected relative to the radiation era, with the additional possibility of significant PBH spin~\cite{Harada:2017fjm,Saito:2023fpt,Saito:2024hlj}.}, for the purposes of the present work, we are considering PBH formation during an initial RD phase which is \emph{followed} by an EMD era. A salient feature of such a period of EMD that we leverage is the possibility of non-negligible accretion for PBHs with masses well below a solar mass $M_{\odot}$~\cite{DeLuca:2021pls}. In the standard accretion scenario that occurs during a period of radiation domination~\cite{Ricotti:2007au,DeLuca:2020bjf}, BHs with masses $M_{\rm PBH}\lesssim M_{\odot}$ are expected to experience negligible accretion. However, during matter domination, substantial accretion (up to orders of magnitude of mass growth) is possible~\cite{1985ApJS...58...39B,DeLuca:2021pls}.

Significant accretion allows for the possibility that the final PBH mass distribution can differ enough from its initial value that the correlation between the SIGW signal from the perturbations that collapsed to form the PBHs and subsequent signals of the final distribution can be altered compared to the expectations in a non-accreting scenario. For example, the SIGW signal spectra produced by the perturbations forming the PBH population is correlated with the PBH masses and their Hawking radiation searches~\cite{Agashe:2022jgk}. Accretion can alter the final form of this relationship, providing scenarios where the PBHs could outgrow their expected doom via Hawking evaporation in their initial mass range, and instead surviving to produce observable Hawking radiation today. Another possibility is that PBHs are produced with a mass that would lead to evaporation at present, but instead have grown to a mass range where such a process is currently negligible.  Such a population may, for example, have shifted into a region where they produce signals in upcoming gravitational lensing observations.

In this paper we explore the effects of an early period of matter domination on  i) SIGWs from perturbations that have produced a PBH population and ii) the alteration of the PBH mass distribution due to non-negligible accretion. We show that upcoming GW observatories such as LISA and BBO could have sensitivity to a PBH population that makes up the entirety of the dark matter in the current universe. Additional signals from this possibility, if detected, could provide insights into aspects of an early era of non-standard cosmology such as its duration and characteristic temperature. Some of our main results include: (1) PBHs formed in the early RD phase could increase their mass through accretion during EMD by up to two orders of magnitude; (2) The resulting GW features a double-peak spectrum, with the high-frequency peak is associated with PBH formation, which is partly suppressed due to the existence of the subsequent EMD, and the low-frequency peak is due to the transition from EMD to late RD phase; (3) If the mass growth of PBHs is of order 10 (100), the mass range where both peaks can be detected by a combination of LISA (BBO) is $10^{15}\ \rm g\lesssim M_{\rm PBH}\lesssim 10^{21}\ \rm g$ $(2\times 10^{15}\ \rm g\lesssim M_{\rm PBH}\lesssim 5\times 10^{18}\ \rm g)$; (4) The signal-to-noise ratio (SNR) of LISA and BBO for the total GW spectrum is calculated for the interesting PBH-as-DM window and its value is generally greater than unity for most cases.

The remainder of this paper is organized as follows. In Sec.~\ref{sec:accretion} we provide the details of the PBH formation mechanism and accretion during a period of early matter domination. In Sec.~\ref{sec:gws} we examine the gravitational wave spectrum produced associated with this PBH population, with results following in Sec.~\ref{sec:results}. We summarize and provide thoughts on future directions in Sec.~\ref{sec:conclusion}. Some supplementary details are given in the Appendices.

\section{Primordial black holes: Formation and accretion}
\label{sec:accretion}

We consider a scenario where PBHs are formed in the RD phase with an initial mass $M_{\rm i}$ that grows due to accretion in a subsequent EMD epoch.

During RD, a horizon-sized perturbation  collapses if the density contrast $\delta$ is greater than the threshold value $\delta_c\approx 0.42$. Assuming a Gaussian distribution of the density contrast, the initial population of PBHs is given by \cite{Carr:1975qj,Allahverdi:2020bys}:
\begin{equation}\label{eq: beta_sigmaH}
    \beta_{\rm i}=\int_{\delta_c}^{O(1)}\frac{1}{\sqrt{2\pi\sigma_{{\rm H}, {\rm i}}^2}}\exp\left(-\frac{\delta^2}{2\sigma_{{\rm H}, {\rm i}}^2}\right)d\delta\approx\rm  Erfc\left[\frac{\delta_c}{\sqrt{2}\sigma_{\rm H, i}}\right],
\end{equation}
where $\rm Erfc$ is the complementary error function and the variance of density contrast $\sigma_{\rm H}(k)\simeq 4\sqrt{\mathcal{P}_\zeta(k)}/9$ is the root mean square amplitude of curvature perturbation at horizon re-entry \cite{Kohri:2018qtx,Harada:2017fjm}\footnote{This simplified relation holds exactly for a scale-invariant spectrum. A more careful treatment involving a choice of the window function is needed for an enhanced power spectrum. Nevertheless, the population and GW spectrum calculated using this approximation \cite{Loc:2024qbz} are almost identical to the results using the window function \cite{Kohri:2024qpd}, which was also noticed in \cite{Kohri:2018qtx}.}. $\beta_{\rm i}$ is defined as the ratio of PBH to radiation energy density at the formation time. We use the subscript ``i" to indicate the re-entry time of modes responsible for the PBH formation. The initial mass of PBHs thus formed is given by:
\begin{equation} \label{mi}
    \left(\frac{M_{\rm i}}{\rm g}\right)=2 \times 10^{31}\left(\frac{\gamma}{0.2}\right) \left({106.75 \over g_{*,{\rm i}}} \right)^{1/2} \left(\frac{\rm GeV}{T_{\rm i}}\right)^2,
\end{equation}
where $\gamma$ is the horizon mass that collapses into the PBH.

Consider an EMD epoch generated by some bosonic field $\phi$, which starts at time $t_{\rm O}$ and ends at time $t_{\rm R}$.
During EMD, PBH starts to accrete in the Bondi-Hoyle regime, resulting in the mass scaling with the scale factor $M \propto a$ \cite{DeLuca:2021pls}. Bondi-Hoyle accretion switches to Eddington accretion at $t_{\rm Edd}$ given by (see Appendix \ref{appendix: Eddington limit})
\begin{equation}\label{eq: t_Edd}
    t_{\rm Edd} \simeq 4\ \frac{H_{\rm O}^{1/5}M_{\rm i}^{6/5}}{\sigma_{{\rm H}, {\rm O}}^{9/5}m_{\rm pl}^{12/5}}.
\end{equation}
where $m_{\rm pl}$ is the Planck mass and $H_{\rm O}$ is the Hubble rate at the onset of EMD. From now on, the subscript ``O" in any quantity denotes its value at the onset of EMD. PBH mass does not change significantly after $t_{\rm Edd}$ because the formation of the disk limits accretion efficiency, the maximum mass will be achieved if the EMD era ends at $t_{\rm R} \gtrsim t_{\rm Edd}$:
\begin{equation}\label{eq: Mmax}
    M_{\rm max}\approx 2 M_{\rm i} \ \sigma_{\rm H,{\rm O}}^{-6/5}\left(\frac{H_{\rm O}}{H_{\rm i}}\right)^{4/5}\approx 2M_{\rm i}\ \sigma_{{\rm H}, {\rm O}}^{-6/5}\left(\frac{T_{\rm O}}{T_{\rm i}}
    \right)^{8/5}.
\end{equation}

The PBH population at the end of EMD follows:
\begin{equation} \label{betar}
    \beta_{\rm R}\approx 2\ \beta_{\rm i}\ \sigma_{{\rm H}, {\rm O}}^{-6/5}\left(\frac{T_{\rm O}}{T_{\rm i}}\right)^{3/5} ,
\end{equation}
where the subscript ``R" denotes the value of a quantity at the end of EMD (onset of RD). Note that we are primarily interested in the PBHs that still survive today, so PBHs must be subdominant at early time before BBN. This means that we still approximately have $\rho_\phi\propto a^{-3}$ despite of accretion. After EMD is over, we have the usual entropy conservation until today. Therefore, the parameter $\beta_{\textnormal{R}}$ can be related to the fraction of the energy density of PBHs to that of DM at present $f_{\rm PBH}$, according to
\begin{equation}
    \beta_{\rm R}=\frac{a_{\rm R}}{a_0}\left(\frac{\Omega_{\rm PBH}}{\Omega_{\rm rad}}\right)_0=\frac{a_{\rm R}}{a_0}\left(\frac{\Omega_{\rm DM}}{\Omega_{\rm rad}}\right)_0 f_{\rm PBH}=\left(\frac{g_{\rm *,0}}{g_{\rm *,R}}\right)^{1/3}\frac{T_0}{T_{\rm R}}\left(\frac{\Omega_{\rm DM}}{\Omega_{\rm rad}}\right)_0 f_{\rm PBH} ,
\end{equation}
where $T_{\rm R}$ is the reheating temperature after EMD and the subscript ``$0$'' denotes the present time. After using the standard values $g_{\rm *,0}=3.94$, $g_{\rm *,R}\simeq 106.75$, $T_0=2.3\times 10^{-13}$ GeV, $\Omega_{\rm DM}=0.27$, and $\Omega_{\rm rad,0}=8.6\times 10^{-5}$, we find
\begin{equation}\label{eq: betaR}
\beta_{\rm R} \approx 2.4\times 10^{-10}\left(\frac{\rm GeV}{T_{\rm R}}\right) f_{\rm PBH}. 
\end{equation}
Then, from Eq.~(\ref{betar}), it follows that:
\begin{equation} \label{eq: fpbh}
    f_{\rm PBH} \approx 8.4\times 10^{9} ~ \beta_{\rm i} ~ \sigma_{{\rm H}, {\rm O}}^{-6/5}\left(\frac{T_{\rm O}}{T_{\rm i}}\right)^{8/5}\left(\frac{T_{\rm R}}{T_{\rm O}}\right)\left(\frac{T_{\rm i}}{\rm GeV}\right) .
\end{equation}
In the limit when there is no EMD $\beta_{\rm R}\rightarrow\beta_i$ and $T_{\rm R}\rightarrow T_i$, Eq. \eqref{eq: betaR} recovers to the standard result (see Eq. (6) of \cite{Carr:2020gox}).

Before moving on to a discussion of GW production, some final comments are in order:
\vskip 2mm
\noindent
(1) Unlike the standard scenario of baryonic accretion, the accretion during EMD is significant regardless of the initial mass of PBHs. This suggests a compelling possibility that small-mass PBHs could accrete to increase their masses and live longer than they would have without accretion by prolonging their time until Hawking evaporation (larger masses evaporate later). Thus, for example, PBHs in the DM window could have originated from smaller PBHs that would have been evaporated by now if there was no EMD era. 
\vskip 2mm
\noindent
(2) As we will show later, PBHs could increase their mass by up to two orders of magnitude. At the same time, the prospective for GW detection in upcoming experiments such as LISA or BBO demands that the final mass should be at least $\sim 10^{14}\rm g$, which means that the initial mass should be at least $10^{12}\rm g$. Even for this lower bound, the lifetime of PBHs after formation should be greater than $1\ \rm s$, which means that these PBHs would not evaporate away before accretion due to EMD becomes effective.

\section{Gravitational Waves}
\label{sec:gws}

In this section, we calculate the GW spectrum of our proposed scenario. There are two contributions: the high-frequency GW associated with PBH formation, which is partly suppressed by the existence of EMD, and the low-frequency GW spectrum produced by the sudden transition from EMD to the late RD phase. For simplicity and consistency, we should also assume a sudden transition from the early RD phase to the EMD phase, so that the modes entering the horizon just before EMD would decay due to the pressure of the plasma. Finally, we compute the SNR for the interesting PBHs-as-DM window.

\subsection{GW generated during RD}

A curvature perturbation $\mathcal{P}_\zeta$ induces tensor modes when the perturbation enters the cosmic horizon \cite{Kohri:2018awv}. The fractional energy density of GW at the production time is 
\begin{equation}\label{eq: Omega GW}
    \Omega_{\rm GW,i}(\eta,k)=\frac{1}{24}\left(\frac{k}{a(\eta)H(\eta)}\right)^2\overline{\mathcal{P}_h(\eta,k)},
\end{equation}
where the oscillation averaged tensor power spectrum is
\begin{equation}
    \overline{P_h(\eta,k)}=2\int_0^\infty dt\int_{-1}^1ds\left[\frac{t(2+t)(s^2-1)}{(1-s+t)(1+s+t)}\right]^2I^2_{\rm RD}(v,u,x)\mathcal{P}_\zeta (kv)\mathcal{P}_\zeta (ku).
\end{equation}
In the deep sub-horizon limit ($x\equiv k \eta =k/(a\,H) \gg 1$), the $\overline{I^2}$ that describes the time evolution of the GW signal due to the transfer function during RD and the gravitational potential source, can be expressed as
\begin{equation}
\begin{aligned}
    \overline{I^2_{\rm RD}(v,u,x\rightarrow\infty)}&=\frac{1}{2}\left[\frac{3(u^2+v^2-3)}{4u^3v^3x}\right]^2\Bigg[\left(-4uv+(u^2+v^2-3)\ln \Bigg|\frac{3-(u+v)^2}{3-(u-v)^2}\Bigg|\right)^2\\
    &+\pi^2(u^2+v^2-3)^2\Theta(v+u-\sqrt{3})\Bigg],
\end{aligned}
\end{equation}
\begin{equation}
    u\equiv\frac{t+s+1}{2};\hspace{1cm} v\equiv\frac{t-s+1}{2}.
\end{equation}
Since $\rho_\phi\propto a^{-3}$ and $\rho_{\rm GW}\propto a^{-4}$, the fractional energy density of GW today is suppressed by a factor $(a_{\rm O}/a_{\rm R})$ relative to the standard result:
\begin{equation}\label{eq: Omega GW RD}
    \Omega_{\rm GW,0}\approx 1.58\times 10^{-4}g_{\rm \rho,i}^{-1/3}\Omega_{\rm GW,i} \left(\frac{a_{\rm O}}{a_{\rm R}}\right),
\end{equation}
where $g_{\rm \rho,i}$ is relativistic degrees of freedom at the emission time.

The frequency $f$ of a GW corresponding to the comoving wavenumber $k$ can be expressed:
\begin{equation}
    \frac{f}{\rm Hz}\approx 1.55\times 10^{-15}\left(\frac{k}{\rm Mpc^{-1}}\right)
\end{equation}
Because of the additional entropy injection during the EMD epoch, one cannot simply use the usual relation between the peak wavenumber associated with PBH formation and PBH mass, which assumes an adiabatic evolution. Then, with the universe expansion during EMD taken into account, we find:
\begin{eqnarray}
k_{\rm p} &=& a_{\rm i} H_{\rm i} \, \nonumber \\
&=&{a_{\rm i} \over a_{\rm O}}{a_{\rm O} \over a_{\rm R}}{a_{\rm R} \over a_{\rm 0}} H_{\rm i}  \, \nonumber \\
&=&\left({T_{\rm O} \over T_{\rm i}}\right)\left({T_{\rm R} \over T_{\rm O}}\right)^{4/3}\left({g_{\rm ,0} \over g_{\rm *,R}}\right)^{1/3}{T_{\rm 0} \over T_{\rm R}}H_{\rm i} \, \nonumber \\
&=&\left({T_{\rm O} \over T_{\rm i}}\right)\left({T_{\rm R} \over T_{\rm O}}\right)^{4/3}\left({g_{\rm *,0} \over g_{\rm *,R}}\right)^{1/3}{T_{\rm 0} \over T_{\rm R}} ~ {\gamma \, m_{\rm pl}^2 \over 2M_{\rm i}} \, .
\end{eqnarray}
This results in
\begin{equation}\label{eq: kp}
    \frac{k_{\rm p}}{{\rm Mpc^{-1}}}  \approx 3.17\times 10^{38}\left({T_{\rm O} \over T_{\rm i}}\right)\left({T_{\rm R} \over T_{\rm O}}\right)^{4/3}\left({{\rm GeV} \over T_{\rm R}}\right)\left({{\rm g} \over M_{\rm i}}\right).
\end{equation}
where we used $\gamma\simeq0.2$. In the limit when there is no EMD $T_{\rm R}\rightarrow T_{\rm O}$, this recovers to the standard result (see Eq. (16) of \cite{Loc:2024qbz}).

\subsection{GW generated at the end of EMD}

The GW signal due to EMD is most observationally prominent when there is a sudden transition from the EMD to the late RD phase \cite{Inomata:2019ivs}. We remain agnostic about the theoretical origin of the EMD and its sudden transition to the RD phase, and take a phenomenological approach, where all we need is the  curvature perturbation amplitude and the duration of the EMD phase. 
The effects of these key entities is that larger perturbations and/or a longer EMD era ({\it i.e.}, an extended perturbation growth period) would create a stronger GW signal. For EMD we have
$\sigma_{\rm H}(k)\simeq (2/5)\sqrt{{\mathcal{P}}_\zeta(k)}$ \cite{Harada:2017fjm}. For example, ${\mathcal{P}}_\zeta \simeq A_s \approx 2 \times 10^{-9}$~\cite{Planck:2018jri} for an (almost) scale-invariant spectrum, as predicted by inflation (assuming the Bunch-Davies vacuum).

We follow \cite{Inomata:2019ivs} to calculate the GW signal associated with the EMD epoch in the sudden-reheating scenario, for different choices of $\sigma_{{\rm H}, {\rm O}}$. The GW spectrum after the reheating transition consists of three contributions from the large-scale modes $\Omega_{\rm GW, RD}^{\rm LS}$, the resonant peak component $\Omega_{\rm GW, RD}^{\rm Res}$, and the GW produced during the EMD epoch $\Omega_{\rm GW,EMD}^{\rm LS}$,
\bea
\Omega_{\rm GW,EMD}=\Omega_{\rm GW, RD}^{\rm LS}+\Omega_{\rm GW, RD}^{\rm Res}+\Omega_{\rm GW, EMD}^{\rm LS}.
\eea
It has been shown that  $\Omega_{\rm GW,EMD}^{\rm LS}$ is subdominant compared to the other two terms~\cite{Inomata:2019ivs}, and is therefore neglected in our calculation. Semi-analytical expressions for the GW spectrum sourced by power-law primordial spectra can be found in Appendix B of \cite{Inomata:2019ivs}.  

To get the maximal GW amplitude (without going beyond the linear regime), we choose the duration of EMD such that the mode that enters the horizon at the beginning of EMD reaches the threshold for non-linear regime at the end of it. For this mode we have
\begin{equation}
    \delta(t)=\sigma_{{\rm H}, {\rm O}} \left(\frac{a(t)}{a_{\rm O}}\right),
\end{equation}
where
\begin{equation}
    \frac{a_{\rm R}}{a_{\rm O}} \simeq \left(\frac{H_{\rm O}}{H_{\rm R}}\right)^{2/3} \approx \left({g_{*,{\rm O}} \over g_{*,{\rm R}}}\right)^{1/3} \left(\frac{T_{\rm O}}{T_{\rm R}}\right)^{4/3}.
\end{equation}
{Setting $\delta(t_{\rm R}) = 1$, we find:} 
\begin{equation} \label{duration}
{T_{\rm R} \over T_{\rm O}} \simeq \left({g_{*,{\rm O}} \over g_{*,{\rm R}}}\right)^{1/4} \sigma_{{\rm H}, {\rm O}}^{3/4}. 
\end{equation}
{Note that for smaller values of $T_{\rm R}/T_{\rm O}$, some modes have entered the non-linear regime by the time the EMD epoch ends.} 

Combining Eqs.~(\ref{eq: fpbh}) and (\ref{duration}), we find the parametric dependence of the current PBH fraction as DM
\begin{equation} \label{fnl}
f_{\rm PBH} \approx 8.4\times 10^9 ~ \beta_{\rm i} ~ \sigma_{{\rm H}, {\rm O}}^{-9/20} \left({g_{*,{\rm O}} \over g_{*,{\rm R}}}\right)^{1/4} \left(\frac{T_{\rm O}}{T_{\rm i}}\right)^{8/5}\left(\frac{T_{\rm i}}{\rm GeV}\right) .
\end{equation}

\section{Results}
\label{sec:results}

After laying out the generic framework, let us now work out an explicit case in detail. We choose the log-normal form, for the enhanced curvature perturbation that created the PBH population, on top of a flat spectrum whose amplitude is parameterized by $A_s$ (as mentioned above):
\begin{eqnarray} \label{lognormal}
    \mathcal{P}_\zeta(k)=\frac{A}{\sqrt{2 \pi \sigma^2}}\exp\left(-\frac{\ln^2(k/k_p)}{2\sigma^2}\right) + A_s ~ ,
\end{eqnarray}
where $A$ is the amplitude, $k_p$ is the peak wavenumber corresponding to PBH formation, and $\sigma$ is the width of the log-normal profile. This form of the power spectrum is typically realized in ultra slow-roll inflationary scenarios \cite{Ballesteros:2020qam,Dalianis:2018frf,Ragavendra:2020sop}. We will choose $\sigma=1$ in this paper as it would induce an almost monochromatic mass function that could be directly compared with observational constraints \cite{Kozaczuk:2021wcl}\footnote{For a discussion on extended PBH mass distributions, see~\cite{Carr:2017jsz} or the reviews~\cite{Carr:2020gox,Carr:2020xqk,Green:2020jor,Auffinger:2022khh}.}.

\begin{figure}[h!]
    \centering
    \includegraphics[width=0.55\linewidth]{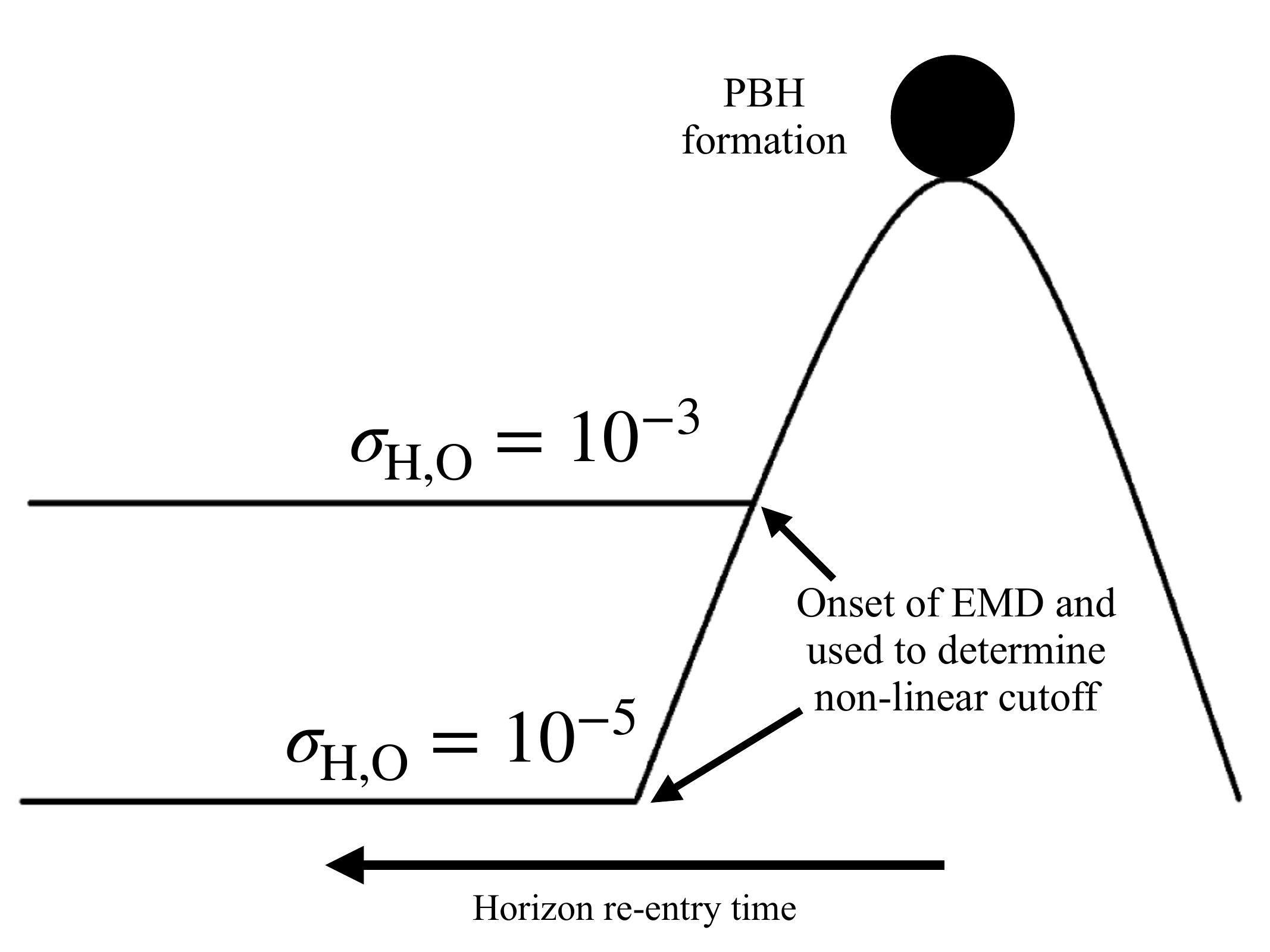}
    \caption{Schematic sketch of the employed curvature power spectrum $\mathcal{P}_\zeta(k)$. PBHs are formed at the peak of the enhanced power spectrum during early RD phase. They then begin to accrete during EMD that begins at either $\sigma_{{\rm H}, {\rm O}}=10^{-3}$ or $\sigma_{{\rm H}, {\rm O}}=10^{-5}$ whose value is also used to determine the nonlinear cutoff (i.e. EMD must be shorter for larger $\sigma_{{\rm H}, {\rm O}}$ in order to not enter the nonlinear regime).}
    \label{fig:sketch}
\end{figure}

{In Fig. \ref{fig:sketch}, we sketch the employed curvature power spectrum $\mathcal{P}_\zeta(k)$. We are considering the case that the onset of EMD is at some point on the low-frequency tail of the enhanced power spectrum. This has several advantages. First, given the robust fact that we need $A\sim O(10^{-2})$ in order to produce PBHs, the correlation between the two peak frequencies given in Eq. \eqref{eq: kokp} offers an opportunity to infer the value of curvature perturbation at the onset of EMD $\sigma_{{\rm H}, {\rm O}}$. In particular, $k_{\rm O}/k_p\sim 10^{-3}\ (10^{-2})$ for $\sigma_{{\rm H}, {\rm O}}=10^{-5}\ (10^{-3})$. Second, from the observational perspective, it is easier to detect both peak frequencies since they are close to each other and could be within the reach of near-future experiments. This will not be the case if $k_{\rm O}/k_p$ is arbitrarily small\footnote{For the first one, the detection of GW peak would not help as an enhanced power spectrum with a short EMD can produce a similar GW spectrum to the case of a smaller power spectrum but with a longer EMD.}.}

{In Appendix \ref{appendix: PBH produced by EMD}, we show that no significant production of PBHs occurs during EMD itself if $\sigma_{{\rm H}, {\rm O}} \lesssim 10^{-3}$, whereas $\sigma_{{\rm H}, {\rm O}}=10^{-5}$ is motivated from the CMB observation. Thus, we focus on the range $10^{-5} \lesssim \sigma_{{\rm H}, {\rm O}}\lesssim 10^{-3}$ in order to avoid a more complicated picture with multiple bouts of PBH production. For illustrative purposes, we also assume a flat $\mathcal{P}_\zeta$ during EMD. The curvature power spectrum could, in principle, take a more complicated form during EMD. Given that our scenario concerns PBH formation in early RD phase and accretion in a subsequent EMD epoch, and accretion only depends on $\mathcal{P}_\zeta$ at the onset of EMD, such complication is unnecessary to consider. In addition, the GW spectrum associated with EMD narrowly peaks at the wavenumber corresponding to the mode entering horizon at the onset of EMD, and hence $\mathcal{P}_\zeta(k_{\rm O})$ is all we need to determine its amplitude.
}

{For the log-normal power spectrum in Eq.~(\ref{lognormal}), the mode that corresponds to $\sigma_{{\rm H}, {\rm O}}$ 
is related to the peak according to:}
\begin{equation}\label{eq: kokp}
    \frac{k_{\rm O}}{k_{ p}}=\exp\left[-\sqrt{-4\sigma^2\ln\left(\frac{5}{2}\frac{(2\pi\sigma^2)^{1/4}}{A^{1/2}}\sigma_{{\rm H}, {\rm O}}\right)}\right].
\end{equation}
{Because the universe is RD from $T_{\rm i}$ down to $T_{\rm O}$, we have $T_{\rm O}/T_{\rm i}=k_{\rm O}/k_{\rm p}$. Therefore, choosing $\sigma=1$ and $A\sim O(10^{-2})$ for the log-normal power spectrum to produce PBHs, the mass growth from Eq.~(\ref{eq: Mmax}) is solely determined by the value of $\sigma_{{\rm H}, {\rm O}}$. As mentioned, we are interested in the $10^{-5}\lesssim\sigma_{{\rm H}, {\rm O}}\lesssim 10^{-3}$ range.}

We present our results focusing on the asteroid-mass window. For a given choice of $\sigma_{{\rm H}, {\rm O}}$ and demanding that $f_{\rm PBH}=1$, we can get all other parameters by using Eqs.~(\ref{eq: beta_sigmaH}), (\ref{eq: Mmax}), (\ref{eq: fpbh}), and~(\ref{eq: kokp}). The results are shown in Tables \ref{tab: sigmaHO 5} and \ref{tab: sigmaHO 3}. We see that for $\sigma_{{\rm H}, {\rm O}}=10^{-5}~(10^{-3})$, the PBH mass can grow by a factor of $\sim 200~(20)$ \footnote{While we mentioned specifically the PBHs-as-DM case, the mass growth is true for other mass ranges as well since PBH population depends exponentially on the amplitude $A$, so we always need $A\sim O(10^{-2})$ in order to have any PBH at all. Thus, the ratio $T_O/T_i$ given in Eq. \eqref{eq: kokp} does not change much, leading to a similar accretion rate in Eq. \eqref{eq: Mmax}.}. 
{The difference in the growth factor can be understood as follows. 
A smaller value of $\sigma_{H, {\rm O}}$ implies a smaller $k_O$ for a fixed $k_p$, see Eq.~(\ref{eq: kokp}), and hence (as explained above) a lower $T_O$. Then Eq.~(\ref{eq: t_Edd}) results in a longer $t_{\rm Edd}${\footnote{The extension of the Eddington time arises because the tangential velocity requires a longer duration to grow and reach the Keplerian velocity necessary for accretion disk formation (see Appendix \ref{appendix: Eddington limit} for details).}, which in turn gives rise to larger growth factor according to Eq.~(\ref{eq: Mmax})}.}

\begin{table}[h!]
    \centering
    \begin{tabular}{|c|c|c|c|c|c|c|}
    \hline
    $M_{\rm i}(\rm g)$ & A & $T_{\rm O}/T_{\rm i}$ & $M_f(\rm g)$ & $T_{\rm R}(\rm GeV)$ & $k_p\rm (Mpc^{-1})$ & $k_{\rm O}\rm (Mpc^{-1})$\\
    \hline
    $10^{15}$ & $2.89\times 10^{-2}$ & $3.07\times 10^{-3}$ & $1.91\times 10^{17}$ & 77.3 & $1.3\times 10^{14}$ & $3.9\times 10^{11}$\\
    \hline
    $10^{17}$ & $3.1\times 10^{-2}$ & $3.04\times 10^{-3}$ & $1.88\times 10^{19}$ & 7.6 & $1.3\times 10^{13}$ & $3.8\times 10^{10}$\\
    \hline
    $10^{20}$ & $3.44\times 10^{-2}$ & $2.98\times 10^{-3}$ & $1.82\times 10^{22}$ & 0.2 & $4.7\times 10^{11}$ & $1.4\times 10^9$\\
    \hline
    \end{tabular}
    \caption{For $\sigma_{{\rm H}, {\rm O}}=10^{-5}$ and $f_{\rm PBH}=1$, we can infer the values of relevant quantities for different choices of PBH initial mass. $M_f$ is the final mass of PBH.}
    \label{tab: sigmaHO 5}
\end{table}

\begin{table}[h!]
    \centering
    \begin{tabular}{|c|c|c|c|c|c|c|}
    \hline
    $M_{\rm i}(\rm g)$ & A & $T_{\rm O}/T_{\rm i}$ & $M_f(\rm g)$ & $T_{\rm R}(\rm GeV)$ & $k_p\rm (Mpc^{-1})$ & $k_{\rm O}\rm (Mpc^{-1})$\\
    \hline
    $10^{16}$ & $2.91\times 10^{-2}$ & $2.1\times 10^{-2}$ & $1.6\times 10^{17}$ & $5194$ &  $1.3\times 10^{14}$ & $2.6\times 10^{12}$\\
    \hline
    $10^{18}$ & $3.11\times 10^{-2}$ &  $2\times 10^{-2}$ & $1.6\times 10^{19}$  & 511 & $1.26\times 10^{13}$ & $2.6\times 10^{11}$\\
    \hline
    $10^{21}$ & $3.47\times 10^{-2}$  & $2\times 10^{-2}$ & $1.5\times 10^{22}$ & 16 & $3.9\times 10^{11}$ & $7.7\times 10^9$\\
    \hline
     \end{tabular}
    \caption{For $\sigma_{{\rm H}, {\rm O}}=10^{-3}$ and $f_{\rm PBH}=1$, we can infer the values of relevant quantities for different choices of PBH initial mass.} 
    \label{tab: sigmaHO 3}
\end{table}

\begin{figure}[h!]
    \centering
    \includegraphics[width=0.75\linewidth]{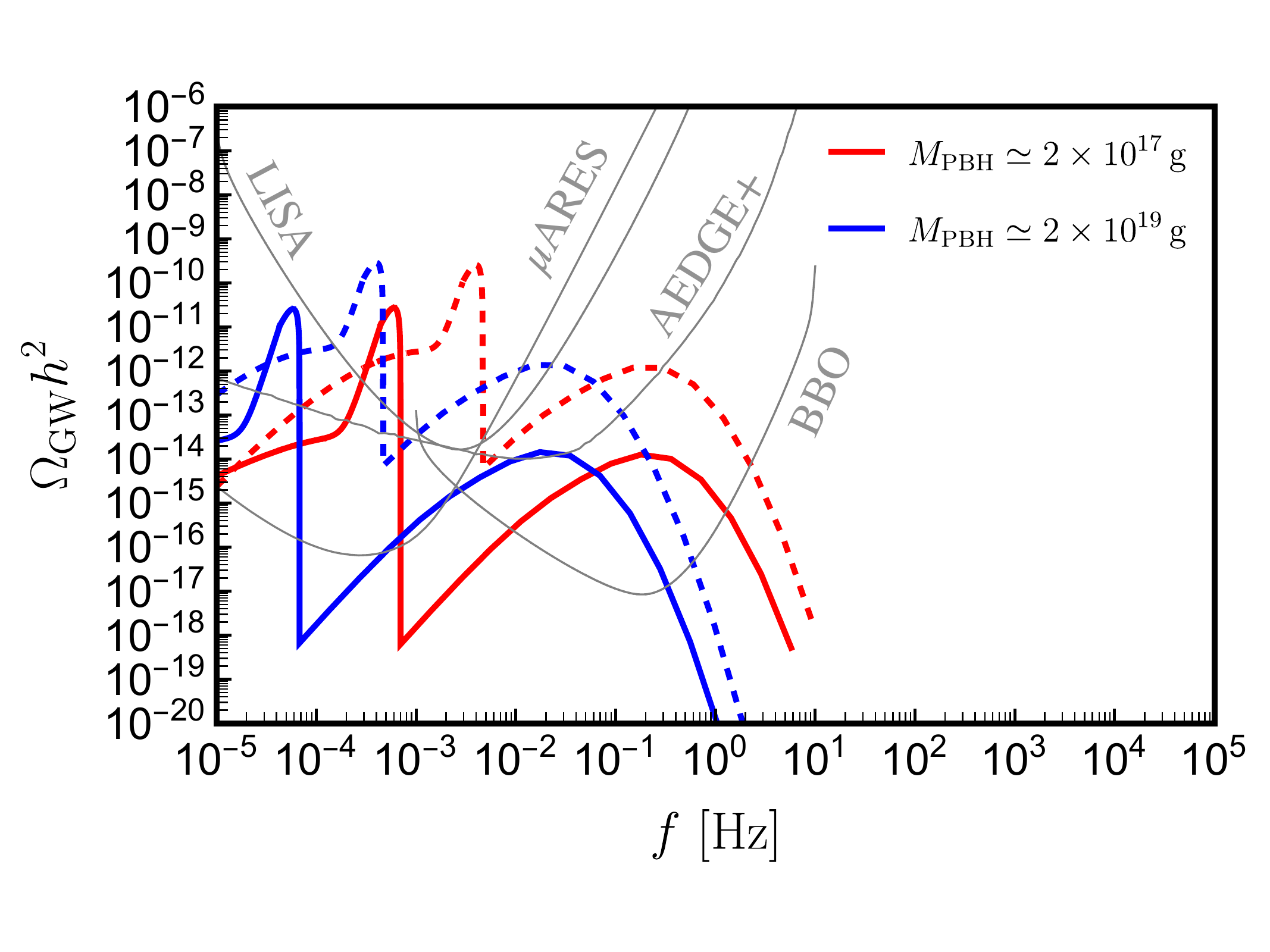}
    \caption{Double-peaked GW signal from benchmark points with different power spectrum amplitude and $M_{\rm PBH}$ after accretion during the EMD epoch. The final PBH mass is $M_{\rm PBH}\simeq2\times10^{17}~{\rm g}$ for the red curves and $M_{\rm PBH}\simeq2\times10^{19}~{\rm g}$ for the blue curves. Solid curves correspond to $\sigma_{{\rm H}, {\rm O}}=10^{-5}$, while dashed curves represent $\sigma_{{\rm H}, {\rm O}}=10^{-3}$.  The projected reach of future GW observatories are shown as grey curves. Sensitivities are obtained from previous works for LISA and BBO~\cite{Schmitz:2020syl}, $\mu$ARES~\cite{Sesana:2019vho}, and AEDGE+~\cite{Badurina:2021rgt}.}
    \label{fig:GW spectrum}
\end{figure}

As an illustration, we show in Fig. \ref{fig:GW spectrum} the GW spectrum for the case $f_{\rm PBH}=1$ with a few choices of PBH mass and the value of curvature perturbation at the onset of EMD (in particular, the first two rows of Tables \ref{tab: sigmaHO 5} and \ref{tab: sigmaHO 3}). The spectrum has two characteristic peaks:
\begin{itemize}
    \item The high-frequency peak is associated with the enhancement of the curvature perturbation that created PBHs during the initial RD phase, which is suppressed partially due to the existence of EMD (see Eq.~\eqref{eq: Omega GW RD}). Increasing $\sigma_{{\rm H}, {\rm O}}$ means shorter EMD, so the amplitude of the GW peak produced during the initial RD phase is less suppressed. The frequency of the RD peaks, however, does not differ significantly when we adjust the value of $\sigma_{{\rm H}, {\rm O}}$. Increasing $\sigma_{{\rm H}, {\rm O}}$ would lead to shorter EMD (smaller $T_{\rm R}/T_{\rm O}$), but it also means the onset of EMD has to be closer to the RD peak (larger $T_{\rm O}/T_{\rm i}$). These two combined effects lead to a similar redshift for the frequency of the RD peaks, as seen in Eq.~(\ref{eq: kp}).
    \item  The low-frequency peak is due to the existence of the EMD phase, which produces a peak at the frequency (or wavenumber) corresponding to the onset of EMD. As shown in \cite{Inomata:2019ivs}, the amplitude of the GW spectrum  produced during EMD depends on the assumed magnitude of the curvature perturbation at the scale corresponding to the onset of EMD. Therefore, increasing $\sigma_{{\rm H}, {\rm O}}$ would lead to an  enhanced GW peak, and the peak frequency is also shifted to the right (higher frequencies) as $k_{\rm O}$ is closer to $k_p$ (see Eq.~(\ref{eq: kokp})).
\end{itemize}

Comparing this GW spectrum with that in the standard RD picture, there are some key factors that differentiate them. First, there is a single GW peak in the case of  the standard cosmological picture of a post-inflationary RD phase without an early EMD era, whereas there are two GW peaks if there was EMD. Second, the amplitude of the GW peak associated with PBH formation is partly suppressed compared to the standard case due to the existence of EMD. For PBHs in the DM window, $\Omega_{\rm GW,peak}^{\rm RD\ only}\sim 10^{-9}$ is set by the PBH formation condition, whereas in our case $\Omega_{\rm GW,peak}^{\rm RD+EMD}\sim 10^{-12}-10^{-14}$ depending on the duration of EMD.

The difference in the peak frequency, however, is minimal. By using the standard result that $k_p^{\rm RD\ only}\simeq 7\times10^{22}(\rm g/M_{\rm i})^{1/2}\ Mpc^{-1}$ \cite{Loc:2024qbz} and Eq. \eqref{eq: kp}, it is easy to show that $k_p^{\rm RD+EMD}/k_P^{\rm RD \ only}\approx \sigma_{{\rm H}, {\rm O}}^{-7/20}(T_{\rm O}/T_{\rm i})^{4/5}\sim0.5$ for $10^{-5}\lesssim\sigma_{{\rm H}, {\rm O}}\lesssim 10^{-3}$. Physically, because of the accretion effect due to EMD, for a given final PBH mass, the initial mass is smaller so that the GW spectrum peaks at a higher frequency. However, the existence of EMD shifts the peak towards a lower frequency that is almost identical to the pure RD case with no accretion.

It can also be seen from Fig. \ref{fig:GW spectrum} that  either one or both peaks could possibly be observed by a combination of different detectors. It is therefore useful to identify the region of the parameter space on the $f_{\rm PBH}-M_{\rm PBH}$ plane where detectability could be achieved. This will help as a useful guide for further developments beyond the scope of our paper, where complementary probes of existing or future Hawking radiation or lensing experiments could test this scenario.

\begin{figure}[h!]
    \centering
    \includegraphics[width=0.49\linewidth]{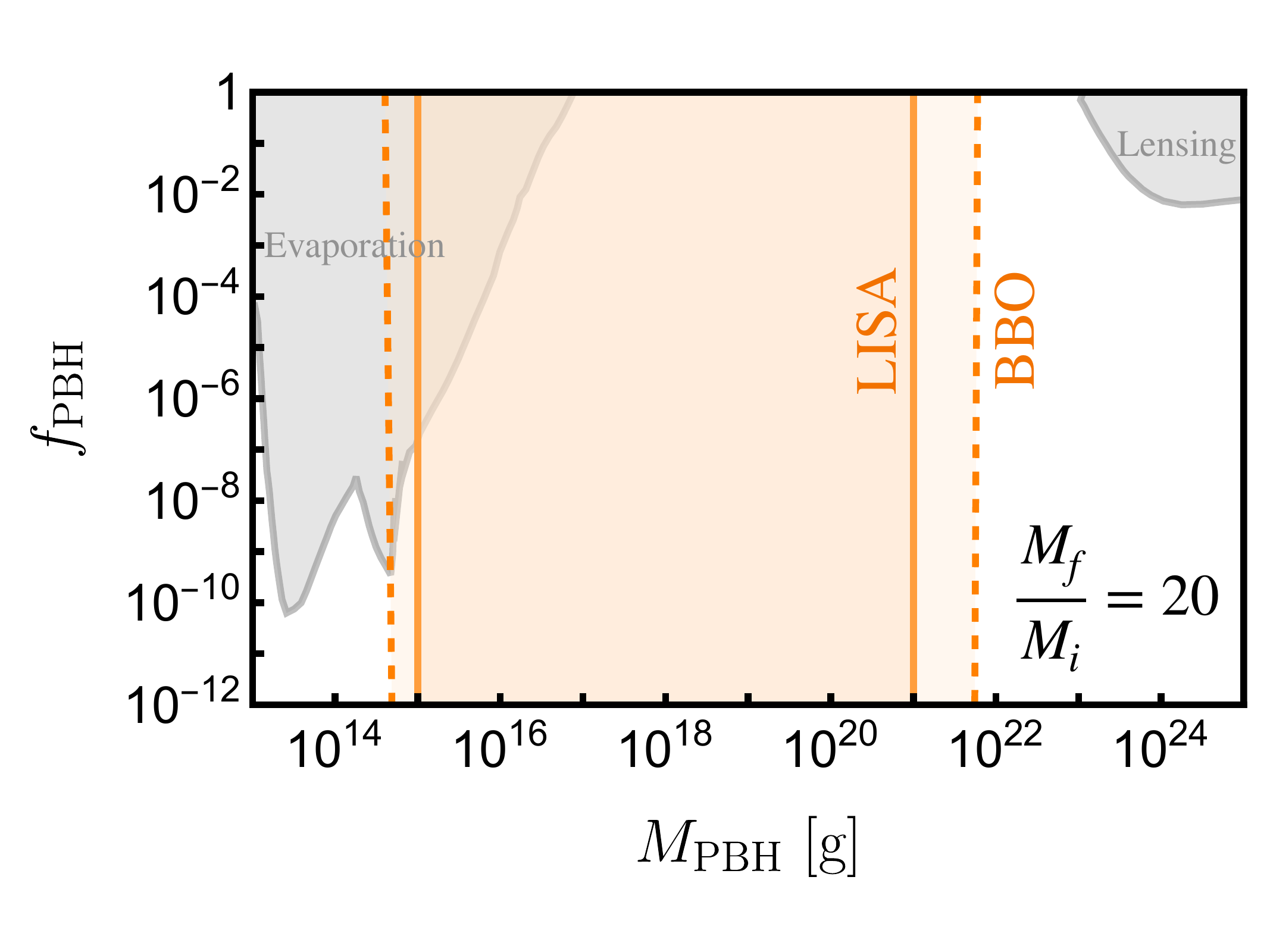}
    \includegraphics[width=0.49\linewidth]{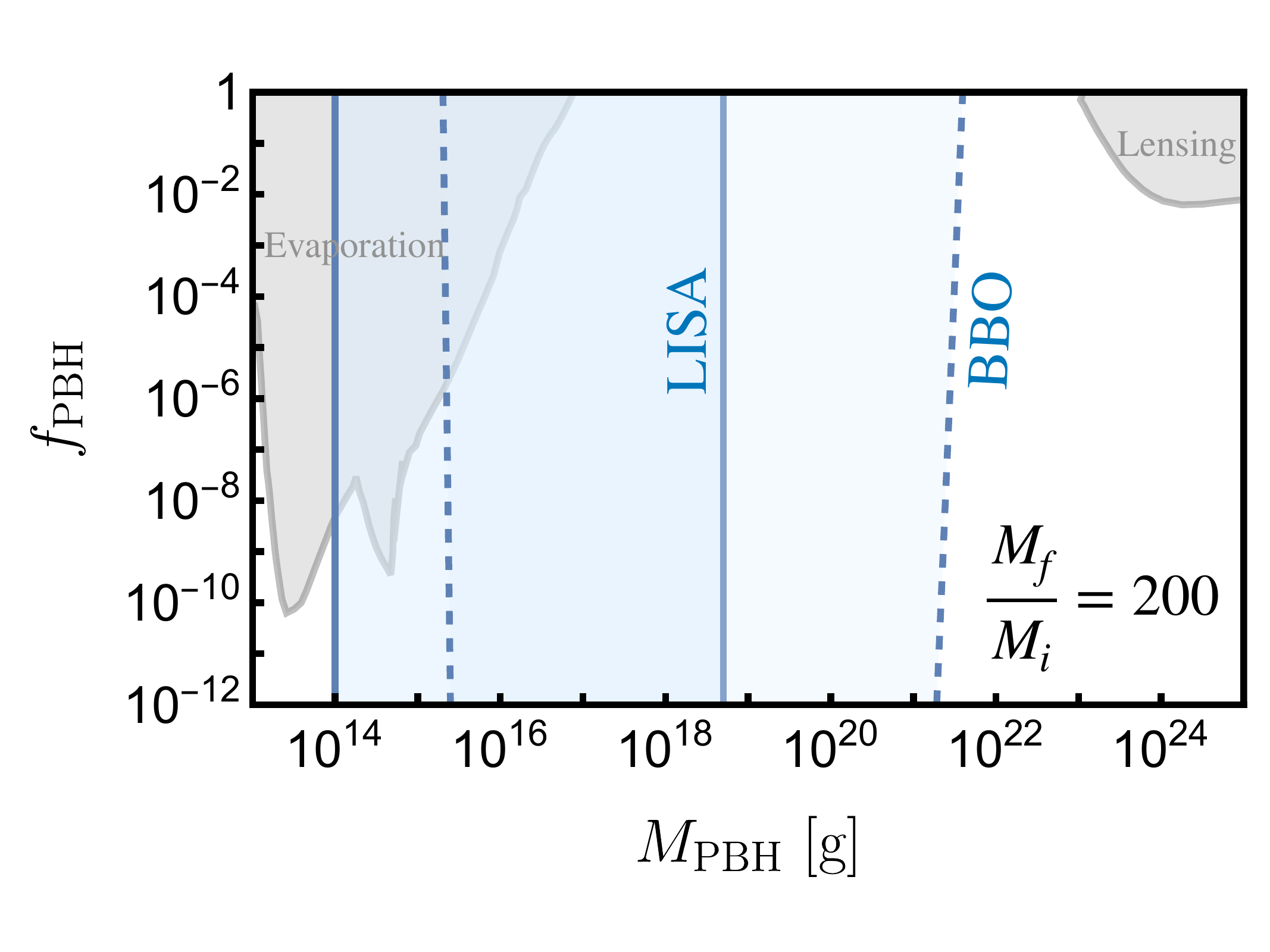}
   
    \caption{PBH parameter space where complementary GW signals at BBO and LISA are expected due to the occurrence of an EMD epoch.
    The regions in between the colored curves can be detected by the named experiments. Solid curves are for the EMD peaks associated with the sudden transition from EMD to late RD phase. Dashed curves are for the RD peaks associated with PBHs' formation. Blue (Orange) color represents $\sigma_{\rm H,O}=10^{-5}\ (10^{-3})$ corresponding to mass growth of $\sim 200\ (20)$. Shaded gray regions are existing constraints from Hawking evaporation and lensing experiments taken from \cite{Carr:2020gox}. }
    \label{fig:fpbh_Mpbh}
\end{figure}

In Fig. \ref{fig:fpbh_Mpbh}, the region between the solid (dashed) curves is where the RD (EMD) peaks could be detected. Note that the BBO (LISA) detector is the most prominent for  detection of the RD (EMD) peaks\footnote{One can also imagine the situation that both peaks can be detected by a single detector. However, it will be fine-tuned and the detectable mass range would be very narrow.}. Also note that the population of PBHs produced during RD depends exponentially on the amplitude of $P_\zeta$ as seen in Eq.~(\ref{eq: beta_sigmaH}), whereas $\Omega_{\rm GW,peak}$ only depends on it quadratically (see Eq.~(\ref{eq: Omega GW})), so that the parameter regions associated with RD peaks appear almost vertically on the $f_{\rm PBH}-M_{\rm PBH}$ plane. The parameter regions associated with the EMD peaks are vertical since the amplitude of GW during EMD is not dictated by the abundance of the PBH population. It is only necessary to track the PBHs population down to $f_{\rm PBH}\sim 10^{-12}$ as PBHs have no physical significance if their population is too small.

Regarding Fig. \ref{fig:fpbh_Mpbh}, some  further comments are in order:
\begin{itemize}
    \item If $\sigma_{{\rm H}, {\rm O}}=10^{-5}$, which corresponds to the maximum mass growth of order 200, the RD peaks can be detected by BBO for the mass range $2 \times10^{15}~\textnormal{g}\lesssim M_f\lesssim 4\times10^{21}~\textnormal{g}$, and the EMD peaks can be detected by LISA for $10^{14}~\textnormal{g}\lesssim M_f\lesssim 5\times 10^{18}~\textnormal{g}$ \footnote{PBHs with mass below $\sim5\times 10^{14}$ g have evaporated by now, so the ``final mass" here is meant to be the mass after accretion.}. Perhaps the most interesting case is when both peaks can be detected, which arises for the range  $2\times10^{15}~\textnormal{g}\lesssim M_f\lesssim  5\times 10^{18}~\textnormal{g}$.
    \item If $\sigma_{{\rm H}, {\rm O}}=10^{-3}$,  which corresponds to a maximum mass growth of roughly a factor of 20, then the RD peaks can be detected by BBO for $4\times 10^{14}~\textnormal{g}\lesssim M_f\lesssim 6\times 10^{21}~\textnormal{g}$, and the EMD peaks can be detected by LISA within $10^{15}~\textnormal{g}\lesssim M_f\lesssim 10^{21}~\textnormal{g}$. The  interesting case where both peaks can be detected arises for $10^{15}~\textnormal{g}\lesssim M_f\lesssim 10^{21}\textnormal{g}$.
    \item Comparing the two cases, the detectable mass ranges of $\sigma_{\rm H,O}=10^{-3}$ are wider than that of $\sigma_{\rm H,O}=10^{-5}$ as the GW spectra (of RD or EMD) move up deep into the detectable regions of GW detectors (see Fig. \ref{fig:GW spectrum}).
\end{itemize}

The EMD must terminate before BBN, which implies $T_{\rm R}\gtrsim 1\ \rm MeV$. By using Eqs. \eqref{duration}, \eqref{eq: kokp}, \eqref{mi} and \eqref{eq: Mmax}, this could be translated into the upper bound on the PBHs' final mass as $M_{\rm PBH}\lesssim [10^{26}~\textnormal{g},10^{30}~\textnormal{g}]$ for $\sigma_{{\rm H}, {\rm O}}=[10^{-5},10^{-3}]$, which corresponds to the initial PBH mass $M_{\rm PBH,i}\lesssim [5\times 10^{23}\ \rm g,5\times 10^{28}\ \rm g]$. This condition is easily satisfied with the above identified mass ranges.

For a wide range of the PBH mass in the asteroid-mass window, the GW signal generated by the enhanced curvature perturbation or the transition from the EMD epoch is much stronger than the detection threshold such that additional information of the spectrum shape can be inferred by correlating across different GW detectors. To show the relative signal strength at GW detectors at different frequencies of our interest, we use LISA and BBO as examples and calculate the signal-to-noise ratio at the two detectors. The SNR is defined as \cite{Schmitz:2020syl}
\begin{equation}
    \rm SNR=\sqrt{t_{\rm obs}\int_{f_{\rm min}}^{f_{\rm max}}df\left(\frac{\Omega_{\rm signal}}{\Omega_{\rm noise}}\right)^2},
\end{equation}
where $t_{\rm obs}$ is the observational time which we take to be one year, the viable frequency range of each detector runs from $f_{\rm min}$ to $f_{\rm max}$, $\Omega_{\rm signal}$ is our combined theoretical GW spectrum of RD and EMD, and $\Omega_{\rm noise}$ is the noise data of each experiment which we took from \cite{Schmitz:2020syl}.

\begin{figure}[h!]
    \centering
    \includegraphics[width=0.65\linewidth]{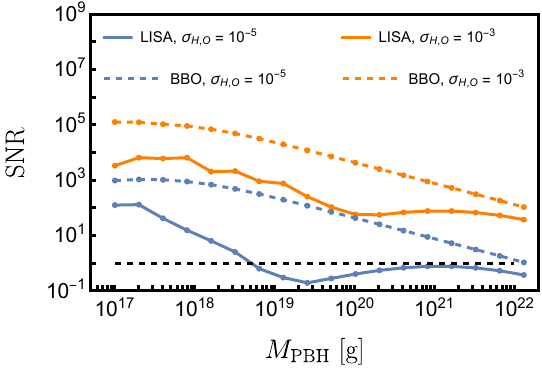}
    \caption{SNR for 1 year of observation as a function PBHs' final mass with $f_{\rm PBH}=1$. The horizontal black dashed line is $\rm SNR=1$ above which the signal is claimed to be detectable.}
    \label{fig:SNR}
\end{figure}

Here we compute the SNR for the double-peaked GW signals focusing on the asteroid-mass window where PBHs could be the totality of DM. PBHs in this mass window are too heavy to have evaporated yet, and are too small to have any significant lensing effects in current searches. Therefore, GW signals associated with their formation in a RD phase and subsequent accretion during EMD presents an intriguing possibility for indirectly inferring the existence of PBHs in this mass window and simultaneously resolving the DM problem.  We present the SNR in Fig. \ref{fig:SNR} for LISA and BBO for a few choices of $\sigma_{{\rm H}, {\rm O}}$. We note that this is the SNR for the \textit{total} GW spectrum which includes the contributions from both the RD and EMD GW spectra. The black horizontal line corresponds to $\rm SNR=1$, above which the signal is claimed to be detectable \footnote{The exact threshold for detectability is still a topic of current discussion, so we present this simply as a guideline.}.

\vspace{0.4cm}

For Fig. \ref{fig:SNR}, we have some additional comments:
\begin{itemize}
    \item The general tendency is that the SNR decreases for increasing PBH mass, as the whole spectrum will move towards lower frequency away from the sensitivity regions. For the SNR of LISA, the EMD peaks move out of LISA's sensitivity region but the RD peaks will move into it, producing a small peak in the SNR at around $M_{\rm PBH}\simeq 2\times10^{21}\ \rm g$ (blue and orange solid curves). For $\sigma_{\rm H,O}=10^{-3}$, SNR of LISA also has a small peak at $M_{\rm PBH}\simeq 2\times 10^{17}\ \rm g$ as GW signal from EMD sits directly above the LISA's noise bottom (cf. Fig. \ref{fig:GW spectrum}).
    \item For a given value of $\sigma_{\rm H,O}$, SNR of BBO is greater than that of LISA as the sensitivity of BBO is much better. Increasing $\sigma_{{\rm H}, {\rm O}}$ results in enhanced SNR as the the RD peak is less  suppressed and the EMD peak is also higher (see Fig. \ref{fig:GW spectrum}). For $\sigma_{{\rm H}, {\rm O}}=10^{-3}$, the SNR of BBO could be as large as $\sim 10^5$ as both RD and EMD peaks coexist inside the sensitivity region and their amplitude is much greater than the noise.
\end{itemize}

We emphasize that Fig. \ref{fig:fpbh_Mpbh} only specifies the mass range where one or both peaks can be detected, which is different from Fig. \ref{fig:SNR} which determines whether or not the total signal could be detected. For instance, for $\sigma_{{\rm H}, {\rm O}}=10^{-5}$, the peaks of GWs associated with PBHs with mass above $\sim 10^{21}~\textnormal{g}$ cannot be detected, but some portion of the GW spectrum still sits above the sensitivity region leading to a SNR that is still greater than unity (dashed blue curve).

\section{Conclusion}
\label{sec:conclusion}

In this paper, we have presented a scenario in which PBHs form in an early post-inflationary radiation dominated (RD) phase, and then accrete during a subsequent early matter dominated (EMD) phase. Depending on the onset time of EMD, the PBH mass could grow by one to two orders of magnitude. We showed that this scenario produces a double-peak GW spectrum, with the high-frequency peak due to the enhanced curvature perturbation required to create PBHs in the early RD phase, and the low-frequency peak arising from the sudden transition from EMD to the late RD phase. We identified the parameter space on the $f_{\rm PBH}-M_{\rm PBH}$ plane   where one can  one or both peaks may be detectable. Additionally, we showed the SNR for GWs produced by PBHs within the asteroid-mass window where PBHs could be the totality of DM.

Looking ahead, there are a few potential further developments beyond the current work. One could consider computing the GW signal when the perturbations have grown into the nonlinear regime, which occurs if EMD lasts long enough. In that case, the GW signal is expected to be stronger than the result presented here \cite{Fernandez:2023ddy}. Another direction is to study how substantial accretion could affect various PBH mass (and spin) distributions, along with the  corresponding observational signatures such as Hawking radiation, gravitational lensing, dynamical effects, or superradiance. These issues are left for future work.

\acknowledgments
The work of R.A. and N.P.D.L. is supported in part by NSF Grant No. PHY-2210367. N.P.D.L. also acknowledges financial support from the UNM Department of Physics and Astronomy through the Origins of the Universe Award. J.B.D. acknowledges support from the National Science Foundation under grant No. PHY-2412995, and thanks the Mitchell Institute at Texas A\&M University for its hospitality where
part of this work was completed. The research activities of T.X. is supported in part by the U.S. National Science Foundation under Grant No. PHY-2412671. We are grateful to the Center for Theoretical Underground Physics and Related Areas (CETUP*), the Institute for Underground Science at Sanford Underground Research Facility (SURF), and the South Dakota Science and Technology Authority for their hospitality and financial support. Their stimulating environment was invaluable during the period in which this work started. 

\appendix

\section{Eddington limit and maximal accretion condition}\label{appendix: Eddington limit}
In this appendix, we review the PBH accretion model with EMD \cite{DeLuca:2021pls}.

At the beginning of EMD, the shell of matter would decouple from the expansion background and collapse if the escape velocity $v_{\rm esc}=\sqrt{2GM_{\rm i}/r_{\rm ita}}$ is smaller than the expansion velocity $v_{\rm expan}=H_{\rm O}r_{\rm ita}$. This gives the initial turn-around radius to be
\begin{equation}
    r_{\rm ita}=\left(\frac{2M_{\rm i}}{m_{\rm pl}^2H_{\rm O}^2}\right)^{1/3}
\end{equation}
Since the density perturbation grows as $\delta\propto a$ during EMD, the PBH mass also grows as $M\propto a$, which implies\footnote{We are assuming that the bosonic field responsible for EMD can be treated as a collisionless fluid, which is only true if its Compton wavelength is much smaller than the PBH's size: 
$$
\lambda_{\rm Compton}\sim\frac{1}{m_\phi}\ll R_S
\Rightarrow m_\phi\gg\frac{m_{\rm pl}^2}{2M_{\rm PBH}}.
$$
This is a rather weak condition. The strongest bound comes from the smallest PBH mass. In our case, we need $m_\phi\gg 10^{-10}\ \rm GeV$ for $M_{\rm PBH}=10^{14}\rm g$ where we used $m_{\rm pl}\sim 10^{19}\ \rm GeV$ and $1\rm g\sim 10^{24}\ \rm GeV$.
}
\begin{equation}\label{eq: Mt}
    M(t)=M_{\rm i}\left(\frac{t}{t_{\rm O}}\right)^{2/3}
\end{equation}
Since $\rho_\phi\propto a^{-3}\propto t^{-2}$ and $M\propto t^{2/3}$, the turn-around radius grows as $r_{\rm ta}\propto (M/\rho_\phi)^{1/3}\propto t^{8/9}$ \cite{Mack:2006gz}. So that 
\begin{equation}
    r_{\rm ta}=r_{\rm ita}\left(\frac{t}{t_{\rm O}}\right)^{8/9}.
\end{equation}
Meanwhile, the velocity perturbation grows as \cite{Kolb:1990vq}
\begin{equation}
    \sigma_v(t)=\sigma_{{\rm H}, {\rm O}}\left(\frac{t}{t_{\rm O}}\right)^{1/3}
\end{equation}

Because of angular momentum conservation, the tangential velocity $v_t$ at radius $r$ is given by: 
\begin{equation}
    v_t\sim\frac{\sigma_vr_{\rm ta}}{r}.
\end{equation}
Whereas the usual Keplerian velocity is given by $v_{\rm Kep}=\sqrt{GM/r}$. The Eddington time $t_{\rm Edd}$ is defined as the time at which an accretion disk forms and the accretion is halted, so the PBH mass does not increase significantly after that. It is determined by
\begin{equation}
    v_t\Bigg|_{r=3r_S}\sim v_{\rm Kep}\Bigg|_{r=3r_S},
\end{equation}
which gives
\begin{equation}
    t_{\rm Edd} \simeq 4\ \frac{H_{\rm O}^{1/5}M_{\rm i}^{6/5}}{\sigma_{{\rm H}, {\rm O}}^{9/5}m_{\rm pl}^{12/5}}.
\end{equation}
Our derivation is the general case when one has the PBH formed already in the prior RD phase and then accretes during EMD. In the limit when PBHs form during EMD (i.e. $T_{\rm O}\rightarrow T_{\rm i}$), one can check that the above equation reduces to the Eq. 30 of \cite{DeLuca:2021pls}. By substituting $t_{\rm Edd}$ into Eq. \eqref{eq: Mt}, we get the maximum mass quoted in Eq. \eqref{eq: Mmax}.

Maximal accretion can only be achieved if $t_{\rm R}\gtrsim t_{\rm Edd}$. By using the above $t_{\rm Edd}$ and $t_R=1/2H_R$ with $H_R^2\propto T_R^4$, this condition is translated into an upper bound on $T_R$:
\begin{equation}
    \left(\frac{T_{\rm R}}{\rm GeV}\right)\lesssim 6.7\times 10^{18}\sigma_{{\rm H}, {\rm O}}^{9/10}\left(\frac{\rm g}{M_{\rm i}}\right)^{3/5}\left(\frac{\rm GeV}{T_{\rm O}}\right)^{1/5}.
\end{equation}
As mentioned in the main text, the GW signal for EMD is most prominent when the EMD lasts long enough. In which case, we can utilize Eq. \eqref{duration} to rewrite this upper bound as
\begin{equation}
    \frac{T_{\rm O}}{T_{\rm i}}\lesssim 1.1\ \sigma_{{\rm H}, {\rm O}}^{1/8}.
\end{equation}
For $10^{-5}\lesssim\sigma_{{\rm H}, {\rm O}}\lesssim 10^{-3}$, we have $T_{\rm O}/T_{\rm i}\lesssim [0.3,0.5]$. On the other hand, Eq. \eqref{eq: kokp} implies that $T_{\rm O}/T_{\rm i}\sim [10^{-3},10^{-2}]$. This means that the maximal accretion condition is much weaker than the nonlinear cutoff condition and can be easily achieved. A partial accretion scenario will therefore need a very short EMD in which the low-frequency GW signal is vanishingly small and the high-frequency GW produced during early RD phase is not suppressed. In that case, one recovers the standard case when there is only one GW peak associated with PBH formation during RD phase. Thus, the maximal accretion scenario that we consider is not only for physical significance but also for distinctive observational signatures.

\section{PBH population produced during EMD}\label{appendix: PBH produced by EMD}
In this appendix, we determine the range of $\sigma_{{\rm H}, {\rm O}}$ such that no significant PBH population can be produced during EMD itself. When taking into account carefully the spin effect with large anisotropy of the collapsing region, which is only significant during EMD, the population of PBH is exponentially suppressed \cite{Harada:2017fjm}

\begin{equation}
    \beta_{\rm i,EMD}\approx 3.244\times 10^{-14}\frac{q^{18}}{\sigma_{{\rm H}, {\rm O}}^4}\exp\left(-0.004608\frac{q^4}{\sigma_{{\rm H}, {\rm O}}^2}\right),
\end{equation}
where $q=1/2$. The final PBH population today is\footnote{We note that PBHs can only form during EMD  when the inward gravitational pull wins over the outward centrifugal force caused by angular momentum effect. Such moment is very close to the nonlinear regime \cite{Harada:2017fjm}. Since we limit ourselves to the linear regime of EMD such that the GW calculation remains valid, PBHs formed during EMD do not experience significant accretion. For the same reason, only modes entering the horizon at the beginning of EMD would eventually form PBHs.}
\begin{equation}
    f_{\rm PBH}^{\rm EMD}\approx 4.2\times 10^9\beta_{\rm i,EMD}\left(\frac{T_{\rm R}}{\rm GeV}\right).
\end{equation}
For $\sigma_{{\rm H}, {\rm O}}=10^{-3}$ and in the most extreme case with $T_{\rm R}=10^{15}\ \rm GeV$, we get $f_{\rm PBH}^{\rm EMD}\sim 10^{-108}$. Thus, only PBHs formed during the early RD phase effectively exist if $\sigma_{{\rm H}, {\rm O}}\lesssim 10^{-3}$. Therefore, we focus on the range $10^{-5}\lesssim\sigma_{{\rm H}, {\rm O}}\lesssim 10^{-3}$ in this paper\footnote{The lower bound is just a suggestion from the CMB observation at large scale and it could in principle take  a smaller value at small scale.}.

\bibliographystyle{JHEP}
\bibliography{reference}

\end{document}